\documentclass[11pt]{article}
\usepackage{amsmath,amssymb}
\usepackage{bm}
\usepackage[dvipdfmx]{graphicx,color}
\usepackage{cite}
\usepackage{geometry}
\DeclareMathOperator{\Tr}{Tr}
\title{Quantum Annealing with Antiferromagnetic Transverse Interactions\\for the Hopfield Model}
\author{Yuya~Seki and Hidetoshi~Nishimori\\
Department of Physics, Tokyo Institute of Technology,\\
Oh-okayama, Meguro-ku, Tokyo 152-8551, Japan}
\date{}
\begin{document}

\maketitle

\begin{abstract}
  We investigate quantum annealing with antiferromagnetic transverse interactions for the generalized Hopfield model with $k$-body interactions.
  The goal is to study the effectiveness of antiferromagnetic interactions, which were shown to help us avoid problematic first-order quantum phase transitions in pure ferromagnetic systems, in random systems.
  We estimate the efficiency of quantum annealing by analyzing phase diagrams for two cases where the number of embedded patterns is finite or extensively large.
  The phase diagrams of the model with finite patterns show that there exist annealing paths that avoid first-order transitions at least for $5 \le k \le 21$.
  The same is true for the extensive case with $k=4$ and $5$.
  In contrast, it is impossible to avoid first-order transitions for the case of finite patterns with $k=3$ and the case of extensive number of patterns with $k=2$ and $3$.
  The spin-glass phase hampers the quantum annealing process in the case of $k=2$ and extensive patterns.
  These results indicate that quantum annealing with antiferromagnetic transverse interactions is efficient also for certain random spin systems.
\end{abstract}

\section{Introduction}
Quantum annealing (QA)~\cite{Kadowaki1998, Kadowaki1998Thesis, Finnila1994, Das2008, Santoro2006, Morita2008, Bapst2013} is a model of quantum computation designed for combinatorial optimization problems.
In particular, adiabatic quantum computation~\cite{Farhi2001} is for the exact solutions of combinatorial optimization problems.
From the physics point of view, some combinatorial optimization problems can be reduced to finding the ground states of Ising spin systems, which is very difficult when the system has a huge number of spin configurations, and the energy landscape is complicated.
A typical example is the ground state search of the three-dimensional spin-glass model.
For this type of problems, QA is known to reach the solution faster than simulated annealing, the classical counterpart, according to numerical~\cite{Kadowaki1998, Kadowaki1998Thesis, Santoro2006} and analytical~\cite{Morita2008} studies although a guaranteed exponential speedup is known only in a single case~\cite{Somma2012}.

Quantum annealing finds the ground state of an Ising spin system according to the following procedure.
First, the Hamiltonian of the Ising model is appended by a term representing quantum fluctuations, typically as a transverse field.
The system is first subjected to a strong transverse field, and the wave function is spread over the whole configuration space under strong quantum fluctuations.
We then reduce quantum fluctuations by controlling the strength of the transverse field, and let the system evolve according to the time-dependent Schr\"{o}dinger equation.
It turns out that an ingenious control of quantum fluctuations drives the wave function toward the ground state of the system.
According to the adiabatic theorem of quantum mechanics~\cite{Messiah1999}, the system stays in the instantaneous ground state during the time evolution if the total time from the initial state, the ground state of the transverse-field only, to the final state, the ground state of the Ising model, is proportional to the inverse square of the minimum energy gap between the instantaneous ground state and the first excited state.
This means that the total computational time grows exponentially fast as a function of the system size if the gap closes exponentially.
It is therefore important to investigate the behavior of the minimum energy gap.

The efficiency of QA in the sense described above is related to statistical-mechanical properties of the system.
According to the finite-size scaling theory, a system with a second-order quantum phase transition has the minimum gap decreasing polynomially with an increasing system size.
This implies that QA can follow the instantaneous ground state and find the desired ground state in a polynomial time.
In contrast, if a system undergoes a first-order quantum phase transition, the gap often decays exponentially at the transition point~\cite{Jorg2008, Jorg2010, Jorg2010:2}, and QA cannot solve the problem efficiently although an anomalous exception is known to exist~\cite{Tsuda2013}.
Thus, we can estimate the efficiency of QA by analyzing the existence and order of quantum phase transitions, which is reflected in the phase diagram.
The phase diagram analysis is more convenient than the energy gap analysis because the phase analysis is not affected by finite-size effects.

The minimum gap may decrease exponentially even without a first-order quantum phase transition.
For example, it has been shown that the minimum gap of the quantum random subcubes model is exponentially small in the system size due to a continuum of level crossing~\cite{Foini2010}.
Although such an exceptional case exists, we focus on the typical case where a first-order quantum phase transition is closely related with difficulties in QA.

Conventional quantum annealing using a transverse field has the following difficulties.
J\"{o}rg \textit{et al}.\ have shown that QA using a transverse field cannot efficiently solve the problem of the simple model with many-body ferromagnetic interactions, whose ground state is the trivial perfect ferromagnet, by showing the existence of a first-order quantum phase transition~\cite{Jorg2010:2}.
Similar arguments have been given in Refs.~\cite{Jorg2008, Jorg2010, Young2010}.

To solve this problem, we have introduced QA using two types of quantum fluctuations induced by a transverse field and antiferromagnetic transverse interactions~\cite{Seki2012} (see also~\cite{Seoane2012}).
We showed that first-order phase transitions in the ferromagnetic model with many-body interactions can be avoided by using antiferromagnetic transverse interactions.
It is interesting to study whether this effectiveness of antiferromagnetic transverse interactions is specific to the ferromagnetic Ising model or it is more generically useful for wider class of models including random cases.
For this purpose, we investigate the Hopfield model with many-body interactions as a typical example of random-spin systems.

The Hopfield model was proposed as a model for associative memory~\cite{Hopfield1982}.
Memories expressed by spin configurations are embedded in the quenched random couplings.
The Hopfield model exhibits different behaviors depending on the number of embedded memory patterns.
If only a single pattern is embedded, the Hopfield model is equivalent to the Mattis model, in which there is no frustration.
This means that the Hopfield model has the same statistical-mechanical properties as the fully connected ferromagnetic model.
In the other extreme limit where the number of embedded patterns is very large, the coupling constants tend to Gaussian variables with zero mean.
This is very similar to the Sherrington-Kirkpatrick (SK) model, although there are still correlations among coupling constants.
We expect that the case with finite patterns greater than one to be an interpolation between the Mattis model and the SK model.
The statistical-mechanical property of the Hopfield model with finite patterns has been investigated by Amit \textit{et al}.~\cite{Amit1985}.
The case of many patterns has been studied in Ref.~\cite{Amit1987}.
Nishimori and Nonomura have developed a full statistical-mechanical analysis of the quantum Hopfield model, i.e., the Hopfield model in a transverse field~\cite{Nishimori1996}.
The statistical-mechanical property of the Hopfield model with many-body interactions has been studied by Gardner~\cite{Gardner1987}.
Ma and Gong have shown the phase diagram of the Hopfield model with many-body interactions in a transverse field in the limit of infinite degree of interactions~\cite{Ma1995}.

The present paper is organized as follows.
Section~\ref{sec: qa} explains QA with antiferromagnetic transverse interactions.
We apply QA with antiferromagnetic transverse interactions to the Hopfield model in Sec.~\ref{sec: application}.
We first show the self-consistent equations, then give the result.
The detailed calculation to derive the self-consistent equations is described in the appendices.
Finally, we conclude in Sec.~\ref{sec: conclusion}.

\section{Quantum annealing and antiferromagnetic transverse interactions}
\label{sec: qa}
We first formulate the procedure of conventional QA.
The system is described by the following time-dependent Hamiltonian:
\begin{align}
 \hat{H}(t)&=s(t)\hat{H}_{0}+[1-s(t)]\hat{V},
\end{align}
where $\hat{H}_{0}$ is the target Hamiltonian whose ground state is to be found.
Considering that $\hat{H}_{0}$ is the Hamiltonian of an Ising spin system, we represent $\hat{H}_{0}$ in terms of the $z$ component of Pauli matrices $\hat{\sigma}_{i}^{z}$ $(i=1,\dotsc , N)$, where $N$ is the number of spins.
The other operator $\hat{V}$ is the driver Hamiltonian, e.g., the transverse-field operator $\hat{V}_{\text{TF}}\equiv -\sum_{i=1}^{N}\hat{\sigma}_{i}^{x}$, that induces quantum fluctuations.
The driver Hamiltonian must not commute with $\hat{H}_{0}$ to induce quantum fluctuations.
We can control quantum fluctuations through $s(t)$.
Since quantum fluctuation is strong at the beginning, we set $s(0)=0$, and fluctuations must eventually vanish, $s(\tau)=1$.
Here, $\tau$ is the running time of QA.

Let us next consider quantum annealing with antiferromagnetic transverse interactions.
This method uses two driver Hamiltonians: One is the following antiferromagnetic interaction
\begin{align}
 \hat{V}_{\!\text{AFF}} &= +N\Bigl(\frac{1}{N}\sum_{i=1}^{N}\hat{\sigma}_{i}^{x}\Bigr)^{2},
 \label{eq:Vaff}
\end{align}
and the other is the conventional transverse-field term $\hat{V}_{\text{TF}}$.
The total Hamiltonian is
\begin{align}
 \hat{H}(s,\lambda ) &= s[\lambda \hat{H}_{0} + (1-\lambda )\hat{V}_{\!\text{AFF}}] + (1-s)\hat{V}_{\text{TF}},\label{eq: total Hamiltonian}
\end{align}
where the control parameters $s$ and $\lambda$ should be changed appropriately as functions of time.
The initial Hamiltonian has $s=0$ and any $\lambda$, and the final Hamiltonian has $s=\lambda =1$.
Intermediate values of $(s,\lambda )$ should be chosen according to the prescription given in the subsequent sections. 
It is convenient to consider the quantum annealing procedure on the $s$-$\lambda$ plane.
A line $\{(s(t),\lambda (t))\mid 0\le t \le \tau\}$ is called an annealing path.
For example, the line $\lambda = 1$ corresponds to the conventional QA since the antiferromagnetic term $\hat{V}_{\!\text{AFF}}$ completely vanishes.
Note that we must keep $\lambda$ strictly positive, however small it is, since quantum fluctuations completely disappear on this line, and the system does not perform quantum annealing processes.

\section{Application to the models}
\label{sec: application}
This section shows the phase diagrams of the Hamiltonian~\eqref{eq: total Hamiltonian}.
We first discuss the Hopfield model with $k$-body interactions and finite patterns embedded.
Next, we study the case with many patterns.

    \subsection{Hopfield model with finite patterns}
    \label{sec: analysis Hopfield finite patterns}
We give self-consistent equations for the Hopfield model with finite patterns embedded.
It is known that the quantum Hopfield model that has two-body interactions exhibits a second-order transition~\cite{Nishimori1996}.
We deal with the case of $k>2$ to check whether antiferromagnetic transverse interactions enable us to avoid a first-order transition.
Furthermore, we restrict the values of $k$ to odd integers in $3 \le k \le 21$ to compare the results with those for the simple ferromagnetic model~\cite{Seki2012}.
Comparing the free energies of symmetric solutions, we show that the phase diagrams are identical with those of the ferromagnetic model with many-body interactions.

        \subsubsection{Self-consistent equations}
The Hamiltonian of the Hopfield model with many-body interactions is given as
\begin{align}
    \hat{H}_{0} &= -\sum_{i_{1}<\dotsb <i_{k}}J_{i_{1},\dotsc ,i_{k}}\hat{\sigma}_{i_{1}}^{z}\dotsm \hat{\sigma}_{i_{k}}^{z}
    \label{eq: Hamiltonian of extension}
\end{align}
with
\begin{align}
    J_{i_{1},\dotsc ,i_{k}} &= \frac{1}{N^{k-1}}\sum_{\mu=1}^{p}\xi_{i_{1}}^{\mu}\dotsm \xi_{i_{k}}^{\mu}.\label{eq: Hebb rule for extension}
\end{align}
Here, $k$ is an integer denoting the degree of interactions, and $\xi_{i}^{\mu}$ takes $\pm 1$ at random.
The number of embedded patterns $p$ is a finite integer independent of $N$.
The total Hamiltonian is given as
\begin{align}
    \hat{H}(s,\lambda) 
    &=
    -s\lambda N \sum_{\mu=1}^{p}\Bigl(\frac{1}{N}\sum_{i=1}^{N}\xi_{i}^{\mu}\hat{\sigma}_{i}^{z}\Bigr)^{k} + s(1-\lambda)N\Bigl(\frac{1}{N}\sum_{i=1}^{N}\hat{\sigma}_{i}^{x}\Bigr)^{2} - (1-s)\sum_{i=1}^{N}\hat{\sigma}_{i}^{x}.\label{eq: many-body interaction Hamiltonian}
\end{align}

We use the mean-field analysis to investigate the phase diagram (see Appendix~\ref{app: finite patterns} for detailed calculations), which gives exact results for the present infinite-range model in the thermodynamic limit.
The order parameters of the Hopfield model are the overlaps with embedded patterns $m_{\mu}$ $(\mu = 1, \dotsc p)$.
In the low-temperature limit $\beta \to \infty $, the pseudo free energy and the self-consistent equations are
\begin{align}
    f(s ,\lambda ; \{m_{\mu}\},m^{x})
    &=
    (k-1)s\lambda\sum_{\mu}(m_{\mu})^{k}-s(1-\lambda)(m^{x})^{2}\notag \\
    &\hphantom{={}}
    -\biggl[\sqrt{\{ks\lambda \sum\nolimits_{\mu}(m_{\mu})^{k-1}\xi^{\mu}\}^{2}+\{1-s-2s(1-\lambda) m^{x}\}^{2}}\,\biggr],\label{eq: free energy in low-temp limit for finite patterns}
\end{align}
and
\begin{align}
    (m_{\mu})^{k-1} &= \left[\frac{ks\lambda \bigl(\sum\nolimits_{\mu}(m_{\mu})^{k-1}\xi^{\mu}\bigr)(m_{\mu})^{k-2}\xi^{\mu}}{\sqrt{\{ks\lambda \sum\nolimits_{\mu}(m_{\mu})^{k-1}\xi^{\mu}\}^{2} + \{1-s-2s(1-\lambda)m^{x}\}^{2}}}\right],\label{eq: hopfield low temp m_mu}\\
    m^{x} &= \left[\frac{1-s-2s(1-\lambda )m^{x}}{\sqrt{\{ks\lambda \sum\nolimits_{\mu}(m_{\mu})^{k-1}\xi^{\mu}\}^{2} + \{1-s-2s(1-\lambda)m^{x}\}^{2}}}\right].\label{eq: hopfield low temp mx}
\end{align}
Here, $m^{x}$ denotes the magnetization along the $x$ direction, and the brackets $[{\dots} ]$ are for the average over the randomness of the embedded patterns.

The self-consistent equations~\eqref{eq: hopfield low temp m_mu} and \eqref{eq: hopfield low temp mx} have the quantum paramagnetic (QP) solution in the region $0 \le s \le 1/(3 - 2\lambda)$.
The order parameters in the QP phase satisfy $m_{\mu}=0$ for all $\mu$ and $m^{x}=1$.
The free energy in the QP phase is
\begin{align}
    f_{\text{QP}}(s, \lambda) &= -s \lambda + 2s -1.
\end{align}

Let us consider the solutions for nonzero $m_{\mu}$'s.
According to the experience in the classical case~\cite{Amit1985}, we expect that the overlaps that give the lowest value of the free energy are symmetric, i.e., $m^{\mu}=m$ for $\mu \le l$ with a given integer $l$, and the others are zero.
The pseudo free energy and self-consistent equations for such symmetric solutions are
\begin{align}
    f_{l}(s,\lambda ; m, m^{x}) &= (k-1)s\lambda l m^{k} -s(1-\lambda )(m^{x})^{2}\notag \\
    &\hphantom{={}}
    - \Bigl[\sqrt{\{ks\lambda m^{k-1}z_{l}\}^{2} + \{1-s-2s(1-\lambda )m^{x}\}^{2}}\,\Bigr],\label{eq: Hopfield free energy zl}
\end{align}
and
\begin{align}
    m &=\left[\frac{ks\lambda m^{k-1}(z_{l})^{2}/l}{\sqrt{\{ks\lambda m^{k-1}z_{l}\}^{2} + \{1-s-2s(1-\lambda )m^{x}\}^{2}}}\right],\label{eq: m zl}\\
    m^{x} &= \left[\frac{1-s-2s(1-\lambda )m^{x}}{\sqrt{\{ks\lambda m^{k-1}z_{l}\}^{2} + \{1-s-2s(1-\lambda )m^{x}\}^{2}}}\right],\label{eq: mx zl}
\end{align}
where we defined the random variable $z_{l} \equiv \sum_{\mu=1}^{l}\xi^{\mu}$.
In particular, for $l=1$, the pseudo free energy $f_{1}$ and the self-consistent equations are identical with those of the many-body interacting ferromagnetic model in the ferromagnetic phase.
This assures us that the phase diagram of the Hopfield model with finite patterns is the same as that of the many-body interacting ferromagnetic model if $f_{1}$ has the lowest value in the symmetric solutions.
The phase for $l=1$ is referred to as the retrieval (R) phase.
The state in the R phase correlates with one of the embedded patterns.

        \subsubsection{Numerical results}
\begin{figure}[tb]
    \centering
    \begin{tabular}[t]{cc}
        \includegraphics[clip]{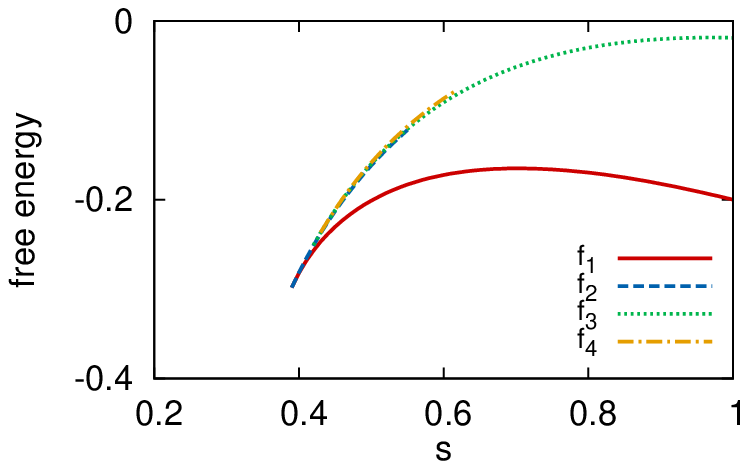} &
        \includegraphics[clip]{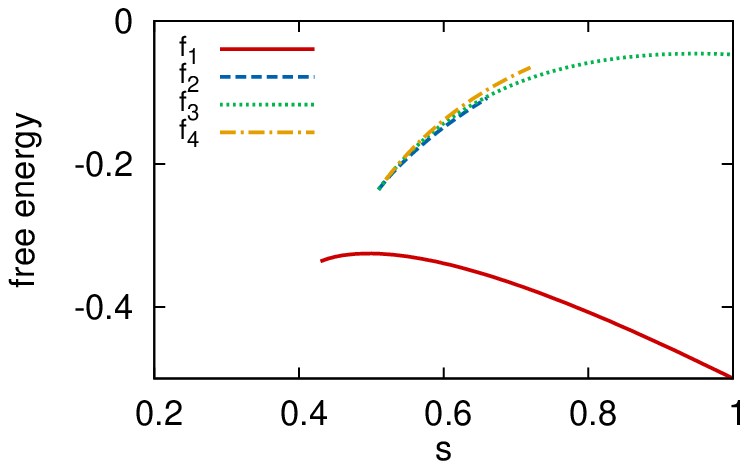} \\
        \includegraphics[clip]{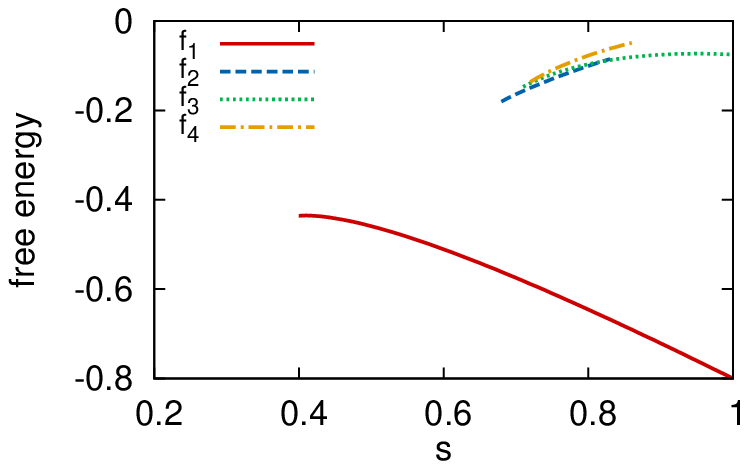} &
        \includegraphics[clip]{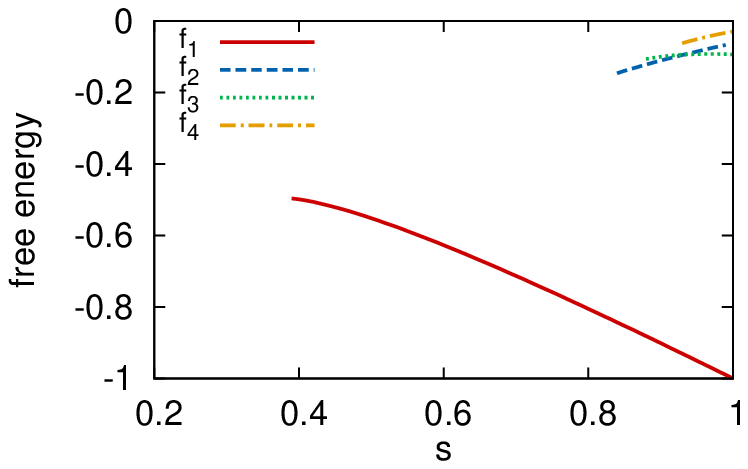} \\
    \end{tabular}
    \caption{(color online) The free energies for symmetric order parameters with $k=5$, $f_{1}$, $f_{2}$, $f_{3}$, and $f_{4}$, for $\lambda = 0.2$ (top left), $0.5$ (top right), $0.8$ (bottom left), and $1$ (bottom right).
       The free energy $f_{1}$ always has the lowest value among the free energies compared.}
    \label{fig: k5}
\end{figure}

We compared the free energies for symmetric order parameters~\eqref{eq: Hopfield free energy zl}, finding that the free energy for the R phase has the lowest value in the free energies among $f_{1}$, $f_{2}$, $f_{3}$, and $f_{4}$, at least for $3 \le k \le 21$ and odd $k$.
We show an example for $k=5$ in Fig.~\ref{fig: k5}.
From this result, we conclude that the R phase is the most stable one among the phases having a symmetric order parameter.

This result indicates that antiferromagnetic transverse interactions greatly improve the process of QA for the generalized Hopfield model with finite patterns.
Since the pseudo free energy and the self-consistent equations for $l=1$ are identical with those of the ferromagnetic model with many-body interactions, the phase diagrams for the generalized Hopfield model with finite patterns are the same as those of the many-body interacting ferromagnetic model shown in Ref.~\cite{Seki2012} except that the ferromagnetic phase is replaced by the R phase.
We have shown in Ref.~\cite{Seki2012} that, whereas the phase transition from the QP phase to the ferromagnetic phase is of first order in the case of three-body interactions, the first-order transition disappears in a range of low $\lambda$ for $5 \le k \le 21$ and odd $k$.
The conventional QA with a transverse field undergoes a first-order quantum phase transition from the QP phase to the R phase.
Antiferromagnetic transverse interactions have thus shown to enable us to avoid the difficulty of QA coming from the first-order phase transitions for $5 \le k \le 21$ and odd $k$, even in the presence of randomness.

    \subsection{Hopfield model with many patterns}
    \label{sec: analysis Hopfield many patterns}
Let us next consider the case of many patterns, i.e., the number of patterns increases as $N$ increases.
Unlike the case of finite patterns, the quantum Hopfield model with $k=2$ exhibits a first-order phase transition between the spin-glass (SG) phase and the R phase.
Hence, we also deal with the $k=2$ case.
First, we analyze the $k=2$ case, and next the case of $k>2$.

        \subsubsection{Self-consistent equations for the case of $k=2$}
The target Hamiltonian is
\begin{align}
    \hat{H}_{0} &= -\frac{1}{2}\sum_{ij}J_{ij}\hat{\sigma}_{i}^{z}\hat{\sigma}_{j}^{z},
\end{align}
where $J_{ij}$ is given as
\begin{align}
    J_{ij}&=\frac{1}{N}\sum_{\mu=1}^{p}\xi_{i}^{\mu}\xi_{j}^{\mu}.
\end{align}
The number of patterns must be proportional to the number of spins $p=\alpha N$ so that the free energy is extensive, as explained in Appendix~\ref{app: Hopfield many patterns}.

To obtain the self-consistent equations, we closely follow Chap.~10 of Ref.~\cite{Hertz1991}.
Detailed calculations are described in Appendix~\ref{app: Hopfield many patterns}.
We assume that the system has a non-vanishing overlap with only one embedded pattern.
Then we have the following self-consistent equations in the low-temperature limit:
\begin{align}
    m &= \int Dz\,\frac{s \lambda m + \sqrt{\alpha \tilde{q}}z}{\sqrt{(s \lambda m + \sqrt{\alpha \tilde{q}}z)^{2}+(1-s-2s(1-\lambda)m^{x})^{2}}},\label{eq: m k=2 low temp}\\
    m^{x} &= \int Dz\,\frac{1-s-2s(1-\lambda)m^{x}}{\sqrt{(s \lambda m + \sqrt{\alpha \tilde{q}}z)^{2}+(1-s-2s(1-\lambda)m^{x})^{2}}},\label{eq: mx k=2 low temp}\\
    q &= \int Dz\,\frac{(s \lambda m + \sqrt{\alpha \tilde{q}}z)^{2}}{(s \lambda m + \sqrt{\alpha \tilde{q}}z)^{2}+(1-s-2s(1-\lambda)m^{x})^{2}},\label{eq: q k=2 low temp}
\end{align}
where $m$ denotes the overlap, $m^{x}$ the magnetization along the $x$ direction, and $q$ the spin-glass order parameter.
We defined the Gaussian measure as $Dz \equiv dz\exp (-z^{2}/2)/\sqrt{2 \pi}$.
The variable $\tilde{q}$ satisfies
\begin{align}
    \tilde{q} &= \frac{(s \lambda)^{2}q}{(1-s \lambda C)^{2}}\label{eq: q tilde k=2 low temp}
\end{align}
with
\begin{align}
    C &= \int Dz\,\frac{\{1-s-2s(1-\lambda)m^{x}\}^{2}}{\{(s \lambda m + \sqrt{\alpha \tilde{q}}z)^{2} + (1-s-2s(1-\lambda)m^{x})^{2}\}^{3/2}}.\label{eq: C k=2 low temp}
\end{align}
The pseudo free energy is written as
\begin{align}
    f &= \frac{1}{2}s \lambda m^{2} - s(1-\lambda)(m^{x})^{2} -\frac{\alpha}{2}s \lambda + \frac{\alpha}{2}\tilde{q}C \notag \\
    &\hphantom{={}}
    -\int Dz\,\sqrt{(s \lambda m + \sqrt{\alpha \tilde{q}})^{2}+(1-s-2s(1-\lambda)m^{x})^{2}}.
\end{align}

        \subsubsection{Phase diagram for the case of $k=2$}
        \label{sec: result k=2}
\begin{figure}[tb]
    \centering
    \includegraphics[clip]{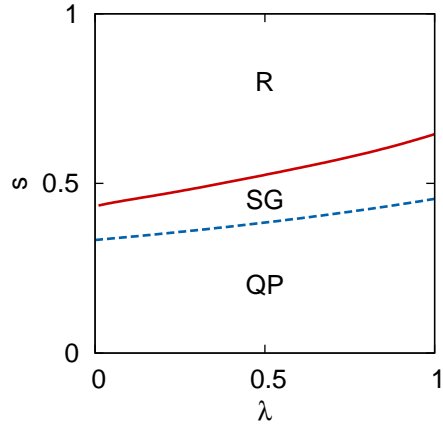}
    \caption{(color online) The phase diagram of the Hopfield model with $k=2$ and with many patterns $p=0.04N$.
    The red solid line represents the first-order phase boundary, and the blue dashed line the second-order boundary.
    The first-order phase transition is inevitable for the QA process.}
    \label{fig: k2alpha004}
\end{figure}

We compared the free energies for three phases: The first is the R phase, $m>0$, the second is the SG phase, $m=0$ and $q>0$, and the last is the QP phase, $m=q=0$.
The phase diagram for the case of $p=0.04N$ is given in Fig.~\ref{fig: k2alpha004}.
Although the phase transition from the QP phase to the SG phase is of second order, the phase transition from the SG phase to the R phase is always of first order.
Therefore, even the method using antiferromagnetic transverse interactions requires an exponentially long time to find the ground state.

The second-order boundary can be obtained analytically.
Expanding Eq.~\eqref{eq: q k=2 low temp} in powers of $q$, we have
\begin{align}
    q&=\alpha\Bigl(\frac{s \lambda}{1-s(3-\lambda)}\Bigr)^{2}q + \mathrm{O}(q^{2}).
\end{align}
Hence, the phase boundary is
\begin{align}
    s&=\frac{1}{3-\lambda (1-\sqrt{\alpha})}\label{eq: phase boundary k=2 QP SG}
\end{align}
or
\begin{align}
    s&=\frac{1}{3-\lambda (1+\sqrt{\alpha})}.\label{eq: spurious phase boundary k=2 QP SG}
\end{align}
Since the boundary~\eqref{eq: phase boundary k=2 QP SG} lies below the other~\eqref{eq: spurious phase boundary k=2 QP SG}, Eq.~\eqref{eq: phase boundary k=2 QP SG} gives the true thermodynamic phase boundary between the QP phase and the SG phase.

        \subsubsection{Self-consistent equations for the case of  $k>2$}
Let us next consider the case of $k>2$.
The Hamiltonian is given by Eqs.~\eqref{eq: Hamiltonian of extension} and \eqref{eq: Hebb rule for extension}.
The number of patterns must be $p=\alpha N^{k-1}$ so that the free energy is extensive.
We consider the case where the system has a non-zero overlap with a single pattern only.
We closely follow the calculation in Ref.~\cite{Gardner1987} to derive the self-consistent equations (see Appendix~\ref{app: Hopfield many patterns many-body} for detailed calculations).
The self-consistent equations in the low-temperature limit are
\begin{align}
    m &= \int Dz\,\frac{s \lambda (km^{k-1} + \sqrt{\alpha kq^{k-1}}z)}{\sqrt{(s \lambda [k m^{k-1} + \sqrt{\alpha k q^{k-1}}z])^{2}+(1-s-2s(1-\lambda)m^{x})^{2}}},\label{eq: m k>2 low temp}\\
    m^{x} &= \int Dz\,\frac{1-s-2s(1-\lambda)m^{x}}{\sqrt{(s \lambda [km^{k-1} + \sqrt{\alpha k q^{k-1}}z])^{2}+(1-s-2s(1-\lambda)m^{x})^{2}}},\label{eq: mx k>2 low temp}\\
    q &= \int Dz\,\frac{(s \lambda [k m^{k-1} + \sqrt{\alpha kq^{k-1}}z])^{2}}{(s \lambda [km^{k-1} + \sqrt{\alpha kq^{k-1}}z])^{2}+(1-s-2s(1-\lambda)m^{x})^{2}}.\label{eq: q k>2 low temp}
\end{align}
The pseudo free energy is
\begin{align}
    f &= s \lambda (k-1) m^{k} - s(1-\lambda)(m^{x})^{2} + \frac{\alpha}{2}k(k-1)(s \lambda)^{2}Cq^{k-1}
    \notag \\
    &\hphantom{={}}
    - \int Dz \sqrt{(s \lambda [k m^{k-1} + \sqrt{\alpha k q^{k-1}}z])^{2}+(1-s-2s(1-\lambda)m^{x})^{2}},
\end{align}
where
\begin{align}
    C = \int Dz\,\frac{\{1-s-2s(1-\lambda)m^{x}\}^{2}}{\{(s \lambda [k m^{k-1} + \sqrt{\alpha k q^{k-1}}z])^{2}+(1-s-2s(1-\lambda)m^{x})^{2}\}^{3/2}}.
\end{align}

        \subsubsection{Phase diagram for the case of $k>2$}
\begin{figure}[tb]
  \centering
    \begin{tabular}{@{}c@{}c@{}c@{}}
      \includegraphics[width=5.5cm, clip]{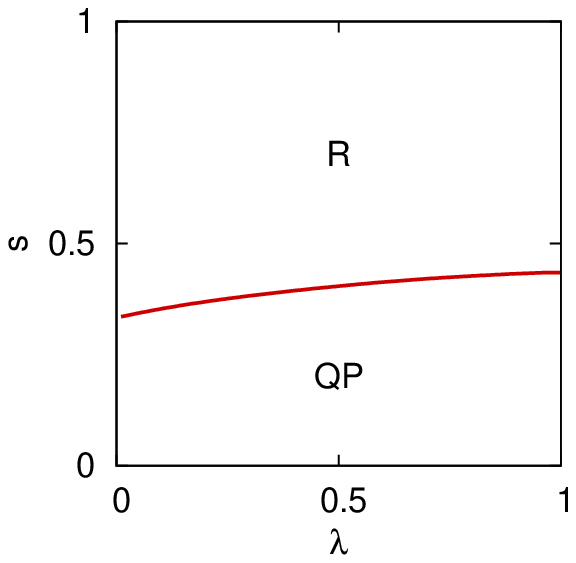}&
      \includegraphics[width=5.5cm, clip]{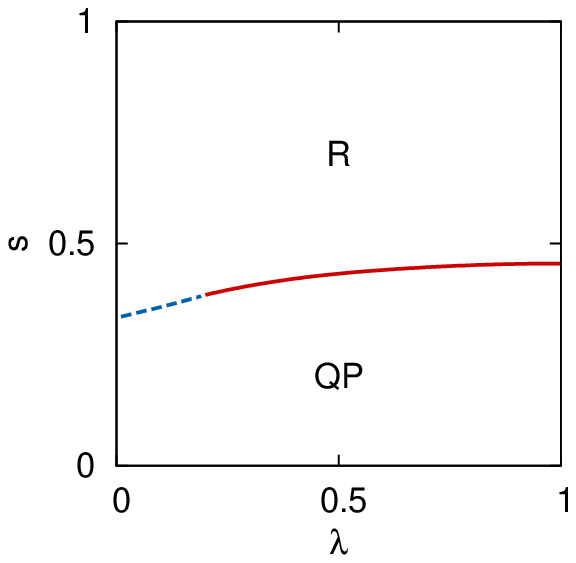}&
      \includegraphics[width=5.5cm, clip]{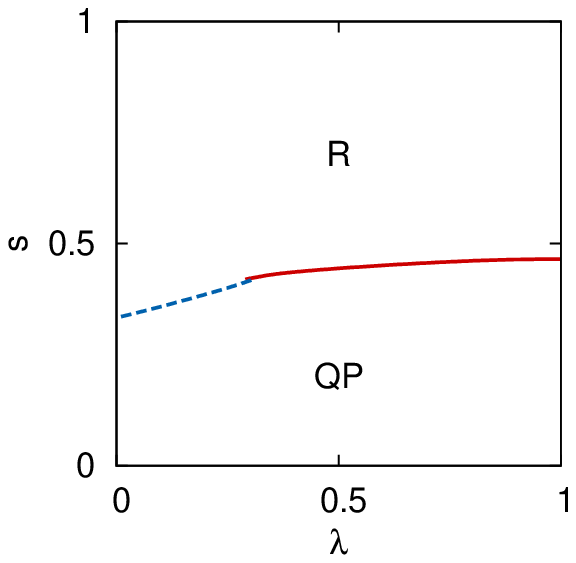}
    \end{tabular}
  
  \caption{(color online) The phase diagram of the generalized Hopfield model with many patterns $p=0.04N^{k-1}$ for $k=3$ (left), $k=4$ (center), $k=5$ (right).
  The red solid line represents the first-order phase boundary, and the blue dashed line the second-order boundary.
  In contrast to the previous case of $k=2$, the SG phase does not appear.
  Whereas the first-order transition is inevitable for $k=3$, we can avoid the first-order transition for $k=4$ and $5$.}
  \label{fig: k>2_alpha004}
\end{figure}

We now show the phase diagram of the generalized Hopfield model with $k=3$, $4$, and $5$ and many patterns $p=0.04N^{k-1}$.
In the same way as in Sec.~\ref{sec: result k=2}, the free energies for the three phases were compared.
We show the resulting phase diagram in Fig.~\ref{fig: k>2_alpha004}.
The SG phase does not appear: The free energy for the SG phase has a higher value than the other free energies for the R phase and the QP phase.
The first-order boundary vanishes for $k=4$ and $5$, and there exist annealing paths to avoid the first-order transition.

\section{Conclusion}
\label{sec: conclusion}
We have studied the effectiveness of antiferromagnetic transverse interactions in QA of the Hopfield model to determine whether or not the success of the application of antiferromagnetic transverse interactions in the ferromagnetic model with many-body interactions is specific to that model.
The analysis of the model is divided into three cases.

First, we have considered the generalized Hopfield model with $k$-body interactions and a finite number of patterns embedded.
The Suzuki-Trotter decomposition and the mean-field analysis have given the self-consistent equations and the pseudo free energy.
We have concluded that the phase diagram is the same as the many-body interacting ferromagnetic model at least for $3 \le k \le 21$ and odd $k$.
Considering the result in Refs.~\cite{Seki2012, Seoane2012}, the present result indicates that antiferromagnetic transverse interactions greatly improve the QA process for the model except for the case of $k=3$.
We conclude that antiferromagnetic transverse interactions are effective also for the random spin system.

Second, the Hopfield model with two-body interactions and extensively many patterns is analyzed.
The difference from the previous case is that the SG phase appears owing to the many unretrieved patterns.
The spins in the SG phase tend to align in the $\pm z$ direction, but do not correlate with any embedded patterns.
We have used the Suzuki-Trotter decomposition, the mean-field analysis, the replica trick, and the static ansatz to study the phase diagram.
The analysis within the RS solution has derived the phase diagram including three phases: the QP phase, the SG phase, and the R phase.
Although the phase boundary between the QP phase and the SG phase is of second order, the boundary between the SG phase and the R phase stays always of first order.
This result indicates difficulties for QA with antiferromagnetic transverse interactions.
Once the system is trapped in a basin in the SG phase, it is hard to escape there to reach the true ground state.

Finally, we have investigated the generalized Hopfield model with many-body interactions and extensively many patterns.
The resulting phase diagram consists of the QP and R phases.
Although the SG solution exists, it has a higher free energy than the other states.
We have confirmed that the first-order phase boundary vanishes at certain values of $\lambda$ for $k=4$ and $k=5$.
Hence, it is possible to avoid the difficulty of exponentially long running time of QA that results from a first-order phase transition.

In conclusion, we have revealed that antiferromagnetic transverse interactions improve the efficiency of QA for some random spin systems.
Using quantum fluctuations other than those induced by a transverse field is helpful for solving combinatorial optimization problems with QA.
In the present paper, we have investigated the efficiency of QA only for the Hopfield model.
It is an interesting problem to identify the class of problems that can be solved by QA with antiferromagnetic transverse interactions.

\section*{Acknowledgment}
Y.S. is grateful for the financial support provided through the Research Fellowship of the Japan Society for the Promotion of Science.

\appendix

\section{Self-consistent equations for the Hopfield model with many-body interactions and finite patterns embedded}
\label{app: finite patterns}
We derive the self-consistent equations~\eqref{eq: hopfield low temp m_mu} and \eqref{eq: hopfield low temp mx} by mean-field analyses.
The Suzuki-Trotter formula and the static ansatz enable us to obtain the partition function.
Then, using a saddle-point condition, we obtain the self-consistent equations.

Let us calculate the partition function.
We first translate the quantum system into a classical system using the Suzuki-Trotter formula~\cite{suzuki1976}.
The Hamiltonian is given as
\begin{align}
 \hat{H}(s,\lambda) &= s\{\lambda \hat{H}_{0} + (1-\lambda)\hat{V}_{\!\text{AFF}}\} + (1-s)\hat{V}_{\text{TF}}\notag \\
 &=
 -s\lambda N \sum_{\mu=1}^{p}\Bigl(\frac{1}{N}\sum_{i=1}^{N}\xi_{i}^{\mu}\hat{\sigma}_{i}^{z}\Bigr)^{k} + s(1-\lambda)N\Bigl(\frac{1}{N}\sum_{i=1}^{N}\hat{\sigma}_{i}^{x}\Bigr)^{2} - (1-s)\sum_{i=1}^{N}\hat{\sigma}_{i}^{x},
\end{align}
where $k$ denotes an integer for the degree of interactions, and $\xi$'s the random variables.
The variable $p$ is an integer independent of $N$.
Using the Trotter decomposition, and introducing $M$ closure relations, we have the following expression of the partition function for a finite Trotter number $M$,
\begin{align}
 Z_{M} &=
 \Tr \prod _{\alpha =1}^{M} \exp \biggl\{\frac{\beta s \lambda N}{M}\sum _{\mu=1}^{p}
 \Bigl(\frac{1}{N}\sum _{i=1}^{N} \xi _{i}^{\mu}\sigma _{i}^{z}(\alpha )\Bigr)^{k}
 - \frac{\beta s(1-\lambda )N}{M}\Bigl(\frac{1}{N}\sum _{i=1}^{N}\sigma _{i}^{x}(\alpha )\Bigr)^{2}\notag \\
&\hphantom{={}}
  + \frac{\beta (1-s)}{M}\sum _{i=1}^{N}\sigma _{i}^{x}(\alpha )\biggr\}
 \prod _{i=1}^{N}\prod_{\alpha =1}^{M}\langle \sigma _{i}^{z}(\alpha )|\sigma _{i}^{x}(\alpha ) \rangle
 \langle \sigma _{i}^{x}(\alpha )|\sigma _{i}^{z}(\alpha + 1) \rangle .
\label{eq: app. p. f. with t. d.}
\end{align}
Here, $\Tr$ denotes the summation over all possible spin configurations of $\{\sigma_{i}^{z}\}$ and $\{\sigma_{i}^{x}\}$ satisfying periodic boundary conditions, $\sigma_{i}^{z}(1)=\sigma_{i}^{z}(M+1)$ for all $i$.
We next linearize the spin-product terms by using delta functions,
\begin{align}
 \delta\Bigl(Nm_{\mu}(\alpha)-\sum_{i}\xi_{i}^{\mu}\sigma_{i}^{z}\Bigr) &= \int d\tilde{m}_{\mu}(\alpha)\,\exp \Bigl\{-\tilde{m}_{\mu}(\alpha)\frac{\beta}{M}\Bigl(Nm_{\mu}(\alpha)-\sum_{i}\xi_{i}^{\mu}\sigma_{i}^{z}(\alpha)\Bigr)\Bigr\},
\end{align}
and
\begin{align}
 \delta\Bigl(Nm^{x}(\alpha)-\sum_{i}\sigma_{i}^{x}\Bigr) &= \int d\tilde{m}^{x}(\alpha)\,\exp \Bigl\{-\tilde{m}^{x}(\alpha)\frac{\beta}{M}\Bigl(Nm^{x}(\alpha)-\sum_{i}\sigma_{i}^{x}(\alpha)\Bigr)\Bigr\}.
\end{align}
Then, Eq.~\eqref{eq: app. p. f. with t. d.} reads
\begin{align}
 Z_{M} &=
 \int \prod _{\alpha ,\mu} dm_{\mu}(\alpha )\,d\tilde{m}_{\mu}(\alpha)\,dm^{x}(\alpha )\,
 d\tilde{m}^{x}(\alpha )\notag \\
 &\hphantom{={}}\times
 \exp \biggl\{N \frac{\beta}{M}\sum _{\alpha}\biggl(s\lambda
 \sum _{\mu}\bigl(m_{\mu}(\alpha)\bigr)^{k}-\sum_{\mu}\tilde{m}_{\mu}(\alpha)m_{\mu}(\alpha) \notag \\
 &\hphantom{={}}\qquad\qquad
 -s(1-\lambda)\bigl(m^{x}(\alpha)\bigr)^{2}
 +(1-s)m^{x}(\alpha)
 -\tilde{m}^{x}(\alpha)m^{x}(\alpha)\biggr)\biggr\}
\notag \\
 &\hphantom{={}}\times
 \exp \biggl\{\sum _{i}\ln \biggl(
 \Tr \exp \frac{\beta}{M}\Bigl\{\sum _{\alpha ,\mu}
 \tilde{m}_{\mu}(\alpha)\xi_{i}^{\mu}\sigma_{i}^{z}(\alpha)
 +\sum _{\alpha}\tilde{m}^{x}(\alpha)\sigma_{i}^{x}(\alpha)\Bigr\} \notag \\
 &\hphantom{={}}\qquad\qquad\times
 \prod_{\alpha} \langle \sigma _{i}^{z}(\alpha) | \sigma _{i}^{x}(\alpha)\rangle
 \langle \sigma_{i}^{x}(\alpha)| \sigma _{i}^{z}(\alpha +1)\rangle \biggr)\biggr\}.\label{eq: app. hopfield Z_M}
\end{align}
In the thermodynamic limit $N\to\infty $, according to the law of large numbers, the summation over the site index $i$ becomes the average over the randomness of the embedded patterns.
We refer to this average as the configurational average.
Furthermore, the integrals are evaluated by the saddle-point method.
The saddle-point conditions for $m_{\mu}(\alpha)$ and $m^{x}(\alpha)$ lead to
\begin{align}
 \tilde{m}_{\mu}(\alpha ) &= s\lambda k\bigl(m_{\mu}(\alpha)\bigr)^{k-1}
\end{align}
and
\begin{align}
 \tilde{m}^{x}(\alpha) &= 1-s-2s(1-\lambda)m^{x}(\alpha),
\end{align}
respectively.
Using the static ansatz, i.e., neglecting the $\alpha$-dependence of the order parameters, we can take the trace in Eq.~\eqref{eq: app. hopfield Z_M} with the inverse operation of the Trotter decomposition.
We thus obtain the following partition function:
\begin{align}
 Z &= \idotsint\prod_{\mu} dm_{\mu}\, dm^{x}\exp\Bigl\{-N\beta\Bigl((k-1)s\lambda\sum_{\mu}(m_{\mu})^{k} -s(1-\lambda)(m^{x})^{2} \notag \\
 &\hphantom{={}}
 -\frac{1}{\beta}\Bigl[\ln 2 \cosh \beta \sqrt{\{ks\lambda \sum\nolimits _{\mu}(m_{\mu})^{k-1}\xi^{\mu}\}^{2}+\{1-s-2s(1-\lambda)m^{x}\}^{2}}\Bigr]\Bigr)\Bigr\},
\end{align}
where the brackets $[{\dots}]$ denote the configurational average.
Therefore the pseudo free energy is
\begin{align}
 &f(\beta ,s ,\lambda ; \{m_{\mu}\},m^{x}) \notag \\
 &=
 (k-1)s\lambda\sum_{\mu}(m_{\mu})^{k}-s(1-\lambda)(m^{x})^{2}\notag \\
 &\hphantom{={}}
 -\frac{1}{\beta}\Bigl[\ln 2 \cosh \beta\sqrt{\{ks\lambda \sum\nolimits_{\mu}(m_{\mu})^{k-1}\xi^{\mu}\}^{2}+\{1-s-2s(1-\lambda) m^{x}\}^{2}}\Bigr],
\end{align}
and the self-consistent equations are
\begin{align}
 (m_{\mu})^{k-1} &= \left[\frac{ks\lambda \bigl(\sum\nolimits_{\mu}(m_{\mu})^{k-1}\xi^{\mu}\bigr)(m_{\mu})^{k-2}\xi^{\mu}}{\sqrt{\{ks\lambda \sum\nolimits_{\mu}(m_{\mu})^{k-1}\xi^{\mu}\}^{2} + \{1-s-2s(1-\lambda)m^{x}\}^{2}}} \right.\notag \\
 &\hphantom{={}}\times \left.
 \vphantom{\frac{ks\lambda \bigl(\sum\nolimits_{\mu}(m_{\mu})^{k-1}\xi^{\mu}\bigr)(m_{\mu})^{k-2}\xi^{\mu}}{\sqrt{\{ks\lambda \sum\nolimits_{\mu}(m_{\mu})^{k-1}\xi^{\mu}\}^{2} + \{1-s-2s(1-\lambda)m^{x}\}^{2}}}}
 \tanh \beta \sqrt{\{ks\lambda \sum\nolimits_{\mu}(m_{\mu})^{k-1}\xi^{\mu}\}^{2} + \{1-s-2s(1-\lambda)m^{x}\}^{2}}\right],
\end{align}
and
\begin{align}
 m^{x} &= \left[\frac{1-s-2s(1-\lambda )m^{x}}{\sqrt{\{ks\lambda \sum\nolimits_{\mu}(m_{\mu})^{k-1}\xi^{\mu}\}^{2} + \{1-s-2s(1-\lambda)m^{x}\}^{2}}} \right.\notag \\
 &\hphantom{={}}\times \left.
 \vphantom{\frac{1-s-2s(1-\lambda )m^{x}}{\sqrt{\{ks\lambda \sum\nolimits_{\mu}(m_{\mu})^{k-1}\xi^{\mu}\}^{2} + \{1-s-2s(1-\lambda)m^{x}\}^{2}}}}
 \tanh \beta \sqrt{\{ks\lambda \sum\nolimits_{\mu}(m_{\mu})^{k-1}\xi^{\mu}\}^{2} + \{1-s-2s(1-\lambda)m^{x}\}^{2}}\right].
\end{align}
In the low-temperature limit $\beta \to \infty $, the pseudo free energy and the self-consistent equations become
\begin{align}
 f(s ,\lambda ; \{m_{\mu}\},m^{x})
 &=
 (k-1)s\lambda\sum_{\mu}(m_{\mu})^{k}-s(1-\lambda)(m^{x})^{2}\notag \\
 &\hphantom{={}}
 -\biggl[\sqrt{\{ks\lambda \sum\nolimits_{\mu}(m_{\mu})^{k-1}\xi^{\mu}\}^{2}+\{1-s-2s(1-\lambda) m^{x}\}^{2}}\,\biggr]
\end{align}
and
\begin{align}
 (m_{\mu})^{k-1} &= \left[\frac{ks\lambda \bigl(\sum\nolimits_{\mu}(m_{\mu})^{k-1}\xi^{\mu}\bigr)(m_{\mu})^{k-2}\xi^{\mu}}{\sqrt{\{ks\lambda \sum\nolimits_{\mu}(m_{\mu})^{k-1}\xi^{\mu}\}^{2} + \{1-s-2s(1-\lambda)m^{x}\}^{2}}}\right],\\
 m^{x} &= \left[\frac{1-s-2s(1-\lambda )m^{x}}{\sqrt{\{ks\lambda \sum\nolimits_{\mu}(m_{\mu})^{k-1}\xi^{\mu}\}^{2} + \{1-s-2s(1-\lambda)m^{x}\}^{2}}}\right].
\end{align}

\section{Self-consistent equations for the Hopfield model with many patterns}
\label{app: Hopfield many patterns}
We derive the self-consistent equations for the Hopfield model with an extensive number of patterns embedded~\eqref{eq: m k=2 low temp}--\eqref{eq: C k=2 low temp}.
We closely follow Chap.~10 of Ref.~\cite{Hertz1991} in the calculation.
The calculation uses the replica trick for configurational average.

Let us calculate the partition function.
In a similar way to the derivation of Eq.~\eqref{eq: app. hopfield Z_M}, the replicated partition function for a Trotter number $M$ is written as
\begin{align}
 [Z_{M}^{n}] &= \int \prod_{\alpha, \mu, \rho} dm_{\mu \rho}(\alpha)\,dm_{\rho}^{x}(\alpha)\, d\tilde{m}_{\rho}^{x}(\alpha) \notag \\
 &\hphantom{={}}\times
 \Tr \exp -\frac{\beta}{M}\sum_{\alpha, \rho}\tilde{m}_{\rho}^{x}(\alpha)\Bigl\{Nm_{\rho}^{x}(\alpha)-\sum_{i}\sigma_{i \rho}^{x}(\alpha)\Bigr\} \notag \\
 &\hphantom{={}}\times
 \exp -\frac{\beta s \lambda N}{2M}\sum_{\alpha, \mu, \rho}\bigl(m_{\mu \rho}(\alpha)\bigr)^{2}
 \times\Bigl[\exp \frac{\beta s \lambda}{M}\sum_{\alpha, \mu, \rho, i} m_{\mu \rho}(\alpha) \xi_{i}^{\mu}\sigma_{i \rho}^{z}(\alpha)\Bigr]  \notag \\
 &\hphantom{={}}\times
 \exp \frac{\beta N}{M}\sum_{\alpha, \rho}\Bigl\{-s(1-\lambda)\bigl(m_{\rho}^{x}(\alpha)\bigr)^{2} + (1-s)m_{\rho}^{x}(\alpha)\Bigr\} \notag \\
 &\hphantom{={}}\times
 \prod_{\alpha, \rho, i}\langle \sigma_{i \rho}^{z}(\alpha) | \sigma_{i \rho}^{x}(\alpha) \rangle \langle \sigma_{i \rho}^{x}(\alpha) | \sigma_{i \rho}^{z}(\alpha + 1) \rangle,
 \label{eq: app. partition function for Hopfield many pattern with M n}
\end{align}
where $\alpha$ $(= 1, \dotsc, M)$ represents the Trotter index, and $\rho$ $(= 1, \dotsc, n)$ the replica index.
We have used a Gaussian integral, instead of the delta function, to linearize the spin-product term regarding $\sigma_{i \rho}^{z}(\alpha)$.

We consider the case where only a single pattern has a non-vanishing overlap with the state of the system: $m_{1 \rho}(\alpha) \equiv m_{\rho}(\alpha)=\mathrm{O}(N^{0})$.
The overlap with the other patterns results from coincidental contributions, hence $m_{\mu \rho}(\alpha) = \mathrm{O}(1/\sqrt{N})$ for $\mu \ge 2$.
Expanding the configurational average for $\mu \ge 2$ in $1/\sqrt{N}$, we have
\begin{align}
 \Bigl[\exp \frac{\beta s \lambda}{M}\sum_{\alpha, \rho, i} m_{\mu \rho}(\alpha)\xi_{i}^{\mu}\sigma_{i \rho}^{z}(\alpha)\Bigr]
 &\simeq
  \exp \frac{\beta ^{2}s^{2}\lambda ^{2}}{2M^{2}} \sum_{i} \sum_{\alpha \rho ,\alpha ' \rho '}m_{\mu \rho }(\alpha )m_{\mu \rho '}(\alpha )\sigma _{i \rho }^{z}(\alpha )\sigma _{i \rho '}^{z}(\alpha ').
\end{align}
Consequently, the term involving $m_{\mu \rho}(\alpha)$ for $\mu \ge 2$ in Eq.~\eqref{eq: app. partition function for Hopfield many pattern with M n} is expressed as a quadratic form:
\begin{align}
 \prod_{\mu \ge 2} \exp -\frac{N \beta s \lambda}{2M}\sum_{\alpha \rho, \alpha' \rho'} \tilde{\Lambda}_{\alpha \rho, \alpha' \rho'}m_{\mu \rho}(\alpha)m_{\mu \rho'}(\alpha'),
\end{align}
with a matrix $\tilde{\Lambda} = (\tilde{\Lambda}_{\alpha \rho, \alpha' \rho'})$,
\begin{align}
 \tilde{\Lambda}_{\alpha \rho, \alpha' \rho'}&\equiv
 \delta_{\alpha \rho, \alpha' \rho'}-\frac{\beta s \lambda}{MN}\sum_{i}\sigma_{i \rho}^{z}(\alpha)\sigma_{i \rho'}^{z}(\alpha').
\end{align}
The integral with regard to $m_{\mu \rho}(\alpha)$ for $\mu \ge 2$ yields
\begin{align}
 (\det \tilde{\Lambda})^{-(p-1)/2} &\simeq (\det \tilde{\Lambda})^{-\alpha N/2}
 =
 \exp -\frac{\alpha N}{2}\ln \det \tilde{\Lambda}
 =
 \exp -\frac{\alpha N}{2}\sum_{\lambda \in \sigma(\tilde{\Lambda})}\ln \lambda,
\end{align}
where we have defined the set of eigenvalues of $\tilde{\Lambda}$ as $\sigma(\tilde{\Lambda})$.
To linearize the spin-product term in $\tilde{\Lambda}$, we replace the matrix by
\begin{align}
\Lambda _{\alpha \rho ,\alpha '\rho '} &\equiv \delta _{\alpha \rho ,\alpha '\rho '}-\frac{\beta s \lambda }{M} q_{\rho \rho '}(\alpha ,\alpha ') -\delta _{\rho \rho '}\frac{\beta s \lambda }{M} R_{\rho }(\alpha ,\alpha ')
\end{align}
with the constraint
\begin{gather}
 q_{\rho \rho '}(\alpha ,\alpha ') = \left\{
 \begin{aligned}
  &\frac{1}{N}\sum_{i} \sigma _{i \rho }^{z}(\alpha )\sigma _{i \rho '}^{z}(\alpha ') && (\rho \ne \rho ')\\
  &0 && (\rho = \rho)
 \end{aligned}
 \right.,\\
 R_{\rho }(\alpha, \alpha') = \frac{1}{N}\sum_{i} \sigma _{i \rho }^{z}(\alpha )\sigma _{i \rho }^{z}(\alpha '),
\end{gather}
introduced by delta functions:
\begin{align}
 &\delta \Bigl(Nq_{\rho \rho'}(\alpha, \alpha') - \sum_{i}\sigma_{i \rho}^{z}(\alpha) \sigma_{i \rho'}^{z}(\alpha')\Bigr) \notag \\
 &=
 \int d \tilde{q}_{\rho \rho'}(\alpha, \alpha') \exp\Bigl\{-\frac{\alpha \beta^{2}}{2M^{2}}\tilde{q}_{\rho \rho'}(\alpha, \alpha')\Bigl(Nq_{\rho \rho'}(\alpha, \alpha')-\sum_{i}\sigma_{i \rho}^{z}(\alpha) \sigma_{i \rho'}^{z}(\alpha')\Bigr)\Bigr\} ,\label{eq: app. delta function for q}\\
 &\delta \Bigl(NR_{\rho}(\alpha, \alpha') - \sum_{i}\sigma_{i \rho}^{z}(\alpha) \sigma_{i \rho}^{z}(\alpha')\Bigr) \notag \\
 &=
 \int d \tilde{R}_{\rho}(\alpha, \alpha') \exp\Bigl\{-\frac{\alpha \beta^{2}}{2M^{2}}\tilde{R}_{\rho}(\alpha, \alpha')\Bigl(NR_{\rho}(\alpha, \alpha')-\sum_{i}\sigma_{i \rho}^{z}(\alpha) \sigma_{i \rho}^{z}(\alpha')\Bigr)\Bigr\}.\label{eq: delta function for R}
\end{align}
Thus, we can rewrite Eq.~\eqref{eq: app. partition function for Hopfield many pattern with M n} as
\begin{align}
 &[Z_{M}^{n}]
 \notag \\
 &= \int \prod_{\alpha \rho } dm_{\rho }(\alpha )\,dm_{\rho }^{x}(\alpha )\,d \tilde{m}_{\rho }^{x}(\alpha ) \prod_{(\alpha \rho ,\alpha '\rho ')} dq_{\rho \rho '}(\alpha ,\alpha ')\, d \tilde{q}_{\rho \rho '}(\alpha ,\alpha ')\prod_{\alpha \alpha' \rho} dR_{\rho }(\alpha ,\alpha ')\, d\tilde{R}_{\rho }(\alpha ,\alpha ')
 \notag \\
 &\hphantom{={}}\times
 \exp \biggl\{-\frac{N \beta s \lambda }{2M}\sum_{\alpha \rho } \bigl(m_{\rho }(\alpha )\bigr)^{2}-\frac{\alpha N}{2}\sum_{\lambda \in \sigma (\Lambda )} \ln \lambda
 \notag \\
 &\hphantom{={}}
 -\frac{N \beta s (1 - \lambda) }{M} \sum_{\alpha \rho } \bigl(m_{\rho }^{x}(\alpha )\bigr)^{2}+\frac{N \beta (1-s)}{M}\sum_{\alpha \rho } m_{\rho }^{x}(\alpha )-\frac{N \beta }{M}\sum_{\alpha \rho } \tilde{m}_{\rho }^{x}(\alpha )m_{\rho }^{x}(\alpha )
 \notag \\
 &\hphantom{={}}
 -\frac{N \alpha \beta^{2}}{2M^{2}}\sum_{(\alpha \rho ,\alpha '\rho ')} \tilde{q}_{\rho \rho '}(\alpha ,\alpha ')q_{\rho \rho '}(\alpha ,\alpha ')- \frac{N \alpha \beta^{2}}{2M^{2}}\sum_{\alpha \alpha' \rho} \tilde{R}_{\rho }(\alpha ,\alpha ')R_{\rho }(\alpha ,\alpha ')
 \biggr\}
 \notag \\
 &\hphantom{={}}\times 
 \biggl[\Tr \exp \biggl\{\frac{\beta s \lambda }{M} \sum_{\alpha \rho } \sum_{i} m_{\rho }(\alpha )\xi _{i}\sigma _{i \rho}^{z}(\alpha )+\frac{\beta }{M} \sum_{\alpha \rho } \sum_{i} \tilde{m}_{\rho }^{x}(\alpha )\sigma _{i \rho }^{x}(\alpha )
 \notag \\
 &\hphantom{={}}
 +\frac{\alpha \beta ^{2}}{2M^{2}} \sum_{(\alpha \rho ,\alpha '\rho ')} \sum_{i} \tilde{q}_{\rho \rho '}(\alpha ,\alpha ')\sigma _{i \rho }^{z}(\alpha )\sigma _{i \rho '}^{z}(\alpha ')+\frac{\alpha \beta ^{2}}{2M^{2}}\sum_{\alpha \alpha' \rho} \sum_{i} \tilde{R}_{\rho }(\alpha ,\alpha ')\sigma _{i \rho }^{z}(\alpha )\sigma _{i \rho}^{z}(\alpha ') \biggr\}
 \notag \\
 &\hphantom{={}}\times 
 \prod_{\alpha \rho i} \langle \sigma _{i \rho }^{z}(\alpha ) | \sigma _{i \rho }^{x}(\alpha ) \rangle
 \langle \sigma _{i \rho }^{x}(\alpha ) | \sigma _{i \rho }^{z}(\alpha +1) \rangle \biggr] .\label{eq: partition function for Hopfield with many patterns linearized}
\end{align}
Here, $\xi_{i}$ denotes $\xi_{i}^{1}$, and $(\alpha \rho, \alpha' \rho')$ all the possible combinations of $\alpha$, $\alpha'$, $\rho$, and $\rho'$ except for the case of $\rho = \rho'$.

We can take the trace in \eqref{eq: partition function for Hopfield with many patterns linearized} independent of $i$.
As a result, Eq.~\eqref{eq: partition function for Hopfield with many patterns linearized} reads
$[Z_{M}^{n}]=\int \prod dm \dots \exp (-N \beta \tilde{f})$, where
\begin{align}
 \tilde{f} &= \frac{s \lambda }{2M} \sum_{\alpha \rho } \bigl(m_{\rho }(\alpha )\bigr)^{2} + \frac{\alpha }{2 \beta }\sum_{\lambda \in \sigma (\Lambda )} \ln \lambda
 \notag \\
 &\hphantom{={}}
 +\frac{s(1-\lambda )}{M} \sum_{\alpha \rho } \bigl(m_{\rho }^{x}(\alpha )\bigr)^{2}-\frac{1-s}{M}\sum_{\alpha \rho } m_{\rho }^{x}(\alpha ) + \frac{1}{M}\sum_{\alpha \rho} \tilde{m}_{\rho }^{x}(\alpha )m_{\rho }^{x}(\alpha )
 \notag \\
 &\hphantom{={}}
 +\frac{\alpha \beta }{2M^{2}} \sum_{(\alpha \rho ,\alpha '\rho ')} \tilde{q}_{\rho \rho '}(\alpha ,\alpha ')q_{\rho \rho '}(\alpha ,\alpha ') + \frac{\alpha \beta }{2M^{2}} \sum_{\alpha \alpha' \rho} \tilde{R}_{\rho }(\alpha ,\alpha ')R_{\rho }(\alpha ,\alpha ')
 \notag \\
 &\hphantom{={}}
 -\frac{1}{\beta }\biggl[\ln \Tr \exp \Bigl\{\frac{\beta s \lambda }{M} \sum_{\alpha \rho } m_{\rho }(\alpha )\sigma _{\rho }^{z}(\alpha )\xi + \frac{\beta }{M} \sum_{\alpha \rho } \tilde{m}_{\rho }^{x}(\alpha )\sigma _{\rho }^{x}(\alpha )
 \notag \\
 &\hphantom{={}}
 +\frac{\alpha \beta ^{2}}{2M^{2}}\sum_{(\alpha \rho ,\alpha '\rho ')} \tilde{q}_{\rho \rho '}(\alpha ,\alpha ')\sigma _{\rho }^{z}(\alpha )\sigma _{\rho '}^{z}(\alpha ') + \frac{\alpha \beta ^{2}}{2M^{2}}\sum_{\alpha \alpha' \rho} \tilde{R}_{\rho }(\alpha ,\alpha ')\sigma _{\rho }^{z}(\alpha )\sigma _{\rho }^{z}(\alpha ') \Bigr\}
 \notag \\
 &\hphantom{={}}\times 
 \prod_{\alpha \rho } \langle \sigma _{\rho }^{z}(\alpha ) | \sigma _{\rho }^{x}(\alpha ) \rangle \langle \sigma _{\rho }^{x}(\alpha ) | \sigma _{\rho }^{z}(\alpha +1) \rangle \biggr].
 \label{eq: f tilde}
\end{align}
The saddle-point conditions for $m_{\rho}(\alpha)$, $\tilde{m}_{\rho}^{x}(\alpha)$, $\tilde{q}_{\rho \rho'}(\alpha, \alpha')$, and $\tilde{R}_{\rho}(\alpha, \alpha')$ lead to the following self-consistent equations:
\begin{align}
 m_{\rho } (\alpha ) &= [\xi \langle \sigma _{\rho }^{z}(\alpha ) \rangle ],\label{eq: S.C.Eq for m before replica trick} \\
 m_{\rho }^{x} (\alpha ) &= [\langle \sigma _{\rho }^{x}(\alpha ) \rangle ], \\
 q_{\rho \rho '} (\alpha ,\alpha ') &= [\langle \sigma _{\rho }^{z}(\alpha )\sigma _{\rho '}^{z}(\alpha ') \rangle ], \\
 R_{\rho} (\alpha ,\alpha ') &= [\langle \sigma _{\rho }^{z}(\alpha )\sigma _{\rho }^{z}(\alpha ') \rangle ],\label{eq: S.C.Eq for R before replica trick}
\end{align}
where the brackets $\langle {\dots} \rangle$ mean the average with respect to the weight
\begin{align}
 &\exp \Bigl\{\frac{\beta s \lambda }{M} \sum_{\alpha \rho } m_{\rho }(\alpha )\sigma _{\rho }^{z}(\alpha )\xi + \frac{\beta }{M} \sum_{\alpha \rho } \tilde{m}_{\rho }^{x}(\alpha )\sigma _{\rho }^{x}(\alpha )
 \notag \\
 &
 +\frac{\alpha \beta ^{2}}{2M^{2}}\sum_{(\alpha \rho ,\alpha '\rho ')} \tilde{q}_{\rho \rho '}(\alpha ,\alpha ')\sigma _{\rho }^{z}(\alpha )\sigma _{\rho '}^{z}(\alpha ') + \frac{\alpha \beta ^{2}}{2M^{2}}\sum_{\alpha \alpha' \rho} \tilde{R}_{\rho }(\alpha ,\alpha ')\sigma _{\rho }^{z}(\alpha )\sigma _{\rho }^{z}(\alpha ') \Bigr\}
 \notag \\
 &\times 
 \prod_{\alpha \rho } \langle \sigma _{\rho }^{z}(\alpha ) | \sigma _{\rho }^{x}(\alpha ) \rangle \langle \sigma _{\rho }^{x}(\alpha ) | \sigma _{\rho }^{z}(\alpha +1) \rangle .
\end{align}

We look for the replica symmetric (RS) solution of Eqs.~\eqref{eq: S.C.Eq for m before replica trick}--\eqref{eq: S.C.Eq for R before replica trick}.
Furthermore, we use the static ansatz, that is, we neglect the dependence of the order parameters on the Trotter number:
\begin{equation}
\begin{alignedat}{3}
 &m_{\rho }(\alpha ) = m, &\quad &m_{\rho }^{x}(\alpha ) = m^{x}, &\quad  &\tilde{m}^{x}_{\rho }(\alpha ) = \tilde{m}^{x},\\
 &q_{\rho \rho '}(\alpha ,\alpha ') = q, &\quad &R_{\rho }(\alpha ,\alpha ') = \left\{\begin{aligned}
    &R && (\alpha \ne \alpha ')\\
    &1 && (\alpha =\alpha )
\end{aligned}\right.,\\
 &\tilde{q}_{\rho \rho '}(\alpha ,\alpha ') = \tilde{q}, &\quad &\tilde{R}_{\rho }(\alpha ,\alpha ') = \tilde{R}.
\end{alignedat}
\end{equation}

First, we evaluate the trace in Eq.~\eqref{eq: f tilde}.
Linearizing the spin-product term by using a Gaussian integral, we can rewrite the term including trace as
\begin{align}
 &n\biggl[ \int Dz\,\ln \Tr \int Dw\,\prod_{\alpha }\exp \Bigl\{\frac{\beta }{M} \Bigl(s \lambda \xi m + \sqrt{\alpha \tilde{q}}z + \sqrt{\alpha (\tilde{R}-\tilde{q})}w\Bigr)\sigma ^{z}(\alpha )\Bigr\}
 \notag \\
 & \times
 \exp \Bigl\{\frac{\beta }{M}\tilde{m}^{x}\sigma ^{x}(\alpha ) \Bigr\} \langle \sigma ^{z}(\alpha ) | \sigma ^{x}(\alpha ) \rangle\langle \sigma ^{x}(\alpha ) | \sigma ^{z}(\alpha + 1 ) \rangle \biggr] + \mathrm{O}(n^{2}),
\end{align}
where $Dz$ denotes the Gaussian measure $Dz \equiv dz\exp (-z^{2}/2)/\sqrt{2 \pi}$, and $Dw$ is defined similarly.
Let us take the limit $M \to \infty$.
Using the inverse operation of the Trotter decomposition, we have
\begin{align}
 &n\biggl[\int Dz\,\ln \int Dw\,2 \cosh \beta \sqrt{\bigl(s \lambda m \xi +\sqrt{\alpha \tilde{q}}z + \sqrt{\alpha (\tilde{R}-\tilde{q})}w\bigr)^{2}+\bigl(\tilde{m}^{x} \bigr)^{2}}\biggr] + \mathrm{O}(n^{2}).\label{eq: Hopfield with many patterns trace term}
\end{align}
The values of the integral are the same for both cases $\xi = 1$ and $\xi = -1$, since the value is invariant under the variable transformation $z \to -z$ and $w \to -w$ .
Hence, Eq.~\eqref{eq: Hopfield with many patterns trace term} reads
\begin{align}
 n\int Dz\,\ln \int Dw\,2 \cosh \beta \sqrt{\bigl(s \lambda m \xi +\sqrt{\alpha \tilde{q}}z + \sqrt{\alpha (\tilde{R}-\tilde{q})}w\bigr)^{2}+\bigl(\tilde{m}^{x} \bigr)^{2}} + \mathrm{O}(n^{2}).
\end{align}

Next, we study the eigenvalues of $\Lambda$.
The matrix has three types of elements:
\begin{align}
\Lambda_{\alpha \rho, \alpha', \rho'} &=\left\{
 \begin{aligned}
  &-\frac{\beta s \lambda}{M}q && \text{if } \rho \ne \rho'\\
  &-\frac{\beta s \lambda}{M}R && \text{if } \rho = \rho' \text{ and } \alpha \ne \alpha'\\
  &1-\frac{\beta s \lambda}{M} && \text{if } \rho = \rho' \text{ and } \alpha = \alpha'
 \end{aligned}
 \right..
\end{align}
We can easily find that the matrix has the eigenvalues:
\begin{align}
 \lambda_{1} &= 1 - \beta s \lambda\Bigl(\frac{1}{M}+\frac{M-1}{M}R+(n-1)q\Bigr)
\end{align}
with degeneracy $1$, and
\begin{align}
 \lambda_{2} &= 1 - \beta s \lambda \Bigl(\frac{1}{M} +\frac{M-1}{M}R - q\Bigr)
\end{align}
with degeneracy $n-1$, and
\begin{align}
\lambda _{3} &= 1- \frac{\beta s \lambda}{M} (1 - R)
\end{align}
with degeneracy $n(M-1)$.
Hence, the eigenvalue sum in Eq.~\eqref{eq: f tilde} reads
\begin{align}
 n\biggl\{\ln (1-\beta s \lambda R + \beta s \lambda q) - \frac{\beta s \lambda q}{1-\beta s \lambda R + \beta s \lambda q}-\beta s \lambda (1-R) \biggr\} + \mathrm{O}(n^{2})
\end{align}

The pseudo free energy is given by using the replica trick:
\begin{align}
 f&=-\frac{1}{N \beta}[\log Z] = -\frac{1}{N \beta}\lim_{n \to 0}\frac{[Z^{n}]-1}{n}
 =
 \lim_{n \to 0}\frac{\tilde{f}}{n}.
\end{align}
From the above results, we obtain
\begin{align}
 f&=\frac{s \lambda }{2}m^{2} + s(1-\lambda )(m^{x})^{2} - (1-s)m^{x} + \tilde{m}^{x}m^{x}
 -\frac{\alpha \beta }{2}\tilde{q}q + \frac{\alpha \beta }{2} \tilde{R}R
 \notag \\
 &\hphantom{={}}
 + \frac{\alpha }{2 \beta } \biggl\{\ln (1-\beta s \lambda R + \beta s \lambda q)-\frac{\beta s \lambda q}{1-\beta s \lambda R + \beta s \lambda q} - \beta s \lambda (1-R) \biggr\}
 \notag \\
 &\hphantom{={}}
 -\frac{1}{\beta } \int Dz\ln \int Dw\,2 \cosh \beta \sqrt{\bigl(s \lambda m +\sqrt{\alpha \tilde{q}}z + \sqrt{\alpha (\tilde{R}-\tilde{q})}w\bigr)^{2}+\bigl(\tilde{m}^{x} \bigr)^{2}}.\label{eq: pseudo free energy Hopfield many patterns k=2}
\end{align}

In what follows, we will derive self-consistent equations in the low-temperature limit.
To simplify expressions shown later, we define the followings:
\begin{align}
 g &\equiv s \lambda m + \sqrt{\alpha \tilde{q}}z + \sqrt{\alpha (\tilde{R}-\tilde{q})}w,\\
 u &\equiv \sqrt{g^{2}+(\tilde{m}^{x})^{2}},\label{eq: def u} \\
 Y &\equiv \int Dw \cosh \beta u.\label{eq: def Y} 
\end{align}
The saddle-point conditions for the pseudo free energy~\eqref{eq: pseudo free energy Hopfield many patterns k=2} leads to the self-consistent equations
\begin{align}
 m &= \int Dz\,Y^{-1}\int Dw\,\frac{g}{u}\sinh \beta u,\label{eq: m k=2}\\
 m^{x} &= \int Dz\,Y^{-1}\int Dw\,\frac{\tilde{m}^{x}}{u}\sinh \beta u, \\
 q &= \int Dz \Bigl(Y^{-1}\int Dw\,\frac{g}{u}\sinh \beta u\Bigr)^{2},\label{eq: q k=2}\\
 R &= \int Dz\,Y^{-1}\Bigl(\int Dw\,\Bigl(\frac{g}{u}\Bigr)^{2}\cosh \beta u + \frac{(\tilde{m}^{x})^{2}}{\beta}\int Dw\,\frac{1}{u^{3}}\sinh \beta u \Bigr)^{2},\label{eq: R k=2}\\
 \tilde{m}^{x} &= 1-s-2s(1-\lambda)m^{x},\label{eq: m tilde k=2} \\
 \tilde{q} &= \frac{(s \lambda)^{2}q}{\{1-\beta s \lambda (R-q)\}^{2}},\label{eq: q tilde k=2}\\
 \tilde{R} &= \tilde{q} + \frac{(s \lambda)^{2}(R-q)}{1-\beta s \lambda (R-q)}.\label{eq: R tilde k=2}
\end{align}
The order parameter $R$ is greater than or equal to $q$, since
\begin{align}
 R&\ge \int Dz\,Y^{-1}\int Dw\,\Bigl(\frac{g}{u}\Bigr)^{2}\cosh \beta u \notag\\
 &=
 \int Dz\,\frac{Y^{-2}}{2 \pi}\int dw\,e^{-w^{2}/2}\cosh \beta u \int dw\,e^{-w^{2}/2}\Bigl(\frac{g}{u}\Bigr)^{2}\cosh \beta u \notag\\
 &\ge
 \int Dz\,\frac{Y^{-2}}{2 \pi}\Bigl\{\int dw (e^{-w^{2}/2}\cosh \beta u)^{1/2}\Bigl(e^{-w^{2}/2}\Bigl(\frac{g}{u}\Bigr)^{2} \cosh \beta u\Bigr)^{1/2}\Bigr\}^{2} \notag\\
 &=
 \int Dz\,\Bigl\{Y^{-1}\int Dw\, \frac{g}{u}\cosh \beta u\Bigr\}^{2} \notag\\
 &\ge
  \int Dz\,\Bigl\{Y^{-1}\int Dw\, \frac{g}{u}\sinh \beta u\Bigr\}^{2} \notag\\
 &=
 q.
\end{align}
In particular, $q$ is equal to $R$ in the limit $\beta \to \infty$ as shown below.
Assuming that $R>q$, we have $\tilde{q}=\tilde{R}=0$ from Eqs.~\eqref{eq: q tilde k=2} and \eqref{eq: R tilde k=2}.
Then, Eqs.~\eqref{eq: q k=2} and \eqref{eq: R k=2} read
\begin{align}
 q&=\Bigl\{\frac{g}{u} \tanh \beta u\Bigr\}^{2} \to \Bigl(\frac{g}{u}\Bigr)^{2},\\
 R&=\Bigl(\frac{g}{u}\Bigr)^{2}+\frac{(\tilde{m}^{x})^{2}\tanh \beta u}{\beta u^{3}} \to \Bigl(\frac{g}{u}\Bigr)^{2},
\end{align}
which is in conflict with the assumption.
Hence, the relation $q=R$ holds in the low-temperature limit.
From Eq.~\eqref{eq: R tilde k=2}, we have $\tilde{q}=\tilde{R}$.
It follows that the integrands in the self-consistent equations are independent of $w$; the integrals with respect to $w$ are taken easily.
Consequently, the self-consistent equations in the low-temperature limit are
\begin{align}
 m &= \int Dz\,\frac{s \lambda m + \sqrt{\alpha \tilde{q}}z}{\sqrt{(s \lambda m + \sqrt{\alpha \tilde{q}}z)^{2}+(1-s-2s(1-\lambda)m^{x})^{2}}},\\
 m^{x} &= \int Dz\,\frac{1-s-2s(1-\lambda)m^{x}}{\sqrt{(s \lambda m + \sqrt{\alpha \tilde{q}}z)^{2}+(1-s-2s(1-\lambda)m^{x})^{2}}},\\
 q &= \int Dz\,\frac{(s \lambda m + \sqrt{\alpha \tilde{q}}z)^{2}}{(s \lambda m + \sqrt{\alpha \tilde{q}}z)^{2}+(1-s-2s(1-\lambda)m^{x})^{2}}.
\end{align}
Although $q$ is equal to $R$, the factor $\beta (R-q)$ converges to
\begin{align}
 \lim_{\beta \to \infty} \beta (R-q) &= \int Dz\,\frac{\{1-s-2s(1-\lambda)m^{x}\}^{2}}{\{(s \lambda m + \sqrt{\alpha \tilde{q}}z)^{2} + (1-s-2s(1-\lambda)m^{x})^{2}\}^{3/2}}
 \equiv C.
\end{align}
For this reason, we obtain
\begin{align}
 \tilde{q} &= \frac{(s \lambda)^{2}q}{(1-s \lambda C)^{2}}.
\end{align}
The pseudo free energy is written as
\begin{align}
 f &= \frac{1}{2}s \lambda m^{2} - s(1-\lambda)(m^{x})^{2} -\frac{\alpha}{2}s \lambda + \frac{\alpha}{2}\tilde{q}C \notag \\
 &\hphantom{={}}
 -\int Dz\,\sqrt{(s \lambda m + \sqrt{\alpha \tilde{q}})^{2}+(1-s-2s(1-\lambda)m^{x})^{2}}.
\end{align}

\section{Self-consistent equations for the Hopfield model with many-body interactions and with many patterns}
\label{app: Hopfield many patterns many-body}
We derive Eqs.~\eqref{eq: m k>2 low temp}--\eqref{eq: q k>2 low temp} in this Appendix.
We closely follow the calculation in Ref.~\cite{Gardner1987}.
The target Hamiltonian is given by Eq.~\eqref{eq: Hamiltonian of extension} and \eqref{eq: Hebb rule for extension}.
The number of patterns must be $p=\alpha N^{k-1}$ so that the free energy is extensive.
We consider the case where the system has a single non-vanishing overlap again.

The replicated partition function for a Trotter number $M$ is calculated in the same way as in the case of $k=2$ except that the spin-product term for $\sigma_{i \rho}^{z}(\alpha)$ is linearized by using the delta function:
\begin{align}
  [Z_{M}^{n}] &= \int \prod_{\alpha, \mu, \rho} dm_{\rho}(\alpha)\,d\tilde{m}_{\rho}(\alpha)\,dm_{\rho}^{x}(\alpha)\, d\tilde{m}_{\rho}^{x}(\alpha) \notag \\
 &\hphantom{={}}\times
 \Tr \Bigl[ \exp -\frac{\beta}{M}\sum_{\alpha, \rho}\tilde{m}_{\rho}(\alpha)\Bigl(Nm_{\rho}(\alpha) - \sum_{i}\xi_{i}^{1}\sigma_{i \rho}^{z}(\alpha)\Bigr)\Bigr] \notag \\
 &\hphantom{={}}\times
 \exp -\frac{\beta}{M}\sum_{\alpha, \rho}\tilde{m}_{\rho}^{x}(\alpha)\Bigl(Nm_{\rho}^{x}(\alpha) - \sum_{i}\sigma_{i \rho}^{x}(\alpha)\Bigr) \notag \\
 &\hphantom{={}}\times
 \exp \frac{\beta s \lambda N}{M}\sum_{\alpha, \rho} \bigl(m_{\rho}(\alpha)\bigr)^{k}
 \prod_{\mu \ge 2} \Bigl[\exp \frac{\beta s \lambda }{MN^{k-1}}\sum_{\alpha, \rho}\sum_{i_{1}< \dotsb <i_{k}}\xi_{i_{1}}^{\mu}\sigma_{i_{1} \rho}^{z}(\alpha)\dotsm \xi_{i_{k}}^{\mu}\sigma_{i_{k} \rho}^{z}(\alpha)
 \Bigr] \notag \\
 &\hphantom{={}}\times
 \exp \frac{\beta N}{M}\sum_{\alpha, \rho}\Bigl\{-s(1-\lambda)\bigl(m_{\rho}^{x}(\alpha)\bigr)^{2} + (1-s)m_{\rho}^{x}(\alpha)\Bigr\} \notag \\
 &\hphantom{={}}\times
 \prod_{\alpha, \rho, i}\langle \sigma_{i \rho}^{z}(\alpha) | \sigma_{i \rho}^{x}(\alpha) \rangle \langle \sigma_{i \rho}^{x}(\alpha) | \sigma_{i \rho}^{z}(\alpha + 1) \rangle .
\end{align}
Note that only the spin-product term for the pattern with non-vanishing overlap is linearized.
The other spin-product term is evaluated as follows.
Expanding the exponential, we find that the linear term in the series vanishes.
The contribution from the second term is
\begin{align}
 &\frac{1}{2}\Bigl(\frac{\beta s \lambda}{MN^{k-1}}\Bigr)^{2}
 \sum_{\alpha,\rho,\alpha',\rho'}\sum_{i_{1}< \dotsb < i_{k}}
 \sigma_{i_{1}\rho}^{z}(\alpha)\sigma_{i_{1}\rho'}^{z}(\alpha')\dotsm
 \sigma_{i_{k}\rho}^{z}(\alpha)\sigma_{i_{k}\rho'}^{z}(\alpha')
 \notag \\
 &=
 \frac{1}{2}\Bigl(\frac{\beta s \lambda}{MN^{k-1}}\Bigr)^{2}
 \sum_{\alpha,\rho,\alpha',\rho'}\Bigl\{\Bigl(\sum_{i}\sigma_{i \rho}^{z}(\alpha)\sigma_{i\rho'}^{z}(\alpha')\Bigr)^{k} + \mathrm{O}(N^{k-1})\Bigr\}
 \notag \\
 &=
 \frac{1}{2}\Bigl(\frac{\beta s \lambda}{M}\Bigr)^{2}\sum_{\alpha,\rho,\alpha',\rho'}
 \frac{1}{N^{k-2}}\Bigl(\frac{1}{N}\sum_{i}\sigma_{i \rho}^{z}(\alpha)\sigma_{i\rho'}^{z}(\alpha')\Bigr)^{k} + \mathrm{O}\Bigl(\frac{1}{N^{k-1}}\Bigr).
\end{align}
Since the contribution from the $l$th term is of the order of $N^{l(1-k/2)}$, the correction to the $\mathrm{O}(N^{2-k})$ term in the series, $\epsilon_{k}$, is the greater one of $\mathrm{O}(N^{1-k})$ and $\mathrm{O}(N^{3(1-k/2)})$: For $k=3$, $\epsilon_{3}=\mathrm{O}(N^{-3/2})$, and for $k>3$, $\epsilon_{k} = \mathrm{O}(N^{1-k})$.
Thus, the series reads
\begin{align}
 &\prod_{\mu \ge 2} \Bigl\{1 + \frac{1}{2}\Bigl(\frac{\beta s \lambda }{M}\Bigr)^{2}\frac{1}{N^{k-2}}\sum_{\alpha, \rho,\alpha',\rho'}\Bigl(\frac{1}{N}\sum_{i}\sigma_{i \rho}^{z}(\alpha)\sigma_{i \rho'}^{z}(\alpha')\Bigr)^{k} + \epsilon_{k}\Bigr\}
 \notag \\
 &=
 \exp \sum_{\mu \ge 2}\Bigl\{\frac{1}{2}\Bigl(\frac{\beta s \lambda}{M}\Bigr)^{2}
 \frac{1}{N^{k-2}}\sum_{\alpha, \rho,\alpha',\rho'}\Bigl(\frac{1}{N}\sum_{i}\sigma_{i \rho}^{z}(\alpha)\sigma_{i \rho'}^{z}(\alpha')\Bigr)^{k} + \epsilon_{k} \Bigr\}
 \notag \\
 &=
 \exp \Bigl\{\frac{\alpha N}{2}\Bigl(\frac{\beta s \lambda}{M}\Bigr)^{2}
 \sum_{\alpha, \rho,\alpha',\rho'}\Bigl(\frac{1}{N}\sum_{i}\sigma_{i \rho}^{z}(\alpha)\sigma_{i \rho'}^{z}(\alpha')\Bigr)^{k} + \alpha N^{k-1}\epsilon_{k}\Bigr\}.\label{eq: higher order terms}
\end{align}
Here, we have used $p= \alpha N^{k-1}$.
The correction term is $\mathrm{O}(N^{1/2})$ for $k=3$, and $\mathrm{O}(N^{0})$ for $k>3$; hence, this term is negligible in the thermodynamic limit.
Linearizing the spin-product term in Eq.~\eqref{eq: higher order terms} by using the delta functions~\eqref{eq: app. delta function for q} and \eqref{eq: delta function for R}, we can write the integrand in $[Z_{M}^{n}]$ as $\exp (-N \beta \tilde{f})$ with
\begin{align}
 \tilde{f}&=-\frac{s \lambda}{M}\sum_{\alpha \rho} \bigl(m_{\rho}(\alpha)\bigr)^{k}
 +\frac{s(1-\lambda)}{M}\sum_{\alpha \rho}\bigl(m_{\rho}^{x}(\alpha)\bigr)^{2}
 -\frac{1-s}{M}\sum_{\alpha \rho}m_{\rho}^{x}(\alpha)\notag \\
 &\hphantom{={}}
 +\frac{1}{M}\sum_{\alpha \rho}\tilde{m}_{\rho}(\alpha)m_{\rho}(\alpha)
 +\frac{1}{M}\sum_{\alpha \rho}\tilde{m}_{\rho}^{x}(\alpha)m_{\rho}^{x}(\alpha)\notag \\
 &\hphantom{={}}
 -\frac{\alpha \beta}{2M^{2}}(s \lambda)^{2}\sum_{(\alpha \rho, \alpha' \rho')}\bigl(q_{\rho \rho'}(\alpha, \alpha')\bigr)^{k}
 -\frac{\alpha \beta}{2M^{2}}(s \lambda)^{2}\sum_{\alpha \alpha' \rho}\bigl(R_{\rho}(\alpha, \alpha')\bigr)^{k}\notag \\
 &\hphantom{={}}
 +\frac{\alpha \beta}{2M^{2}}\sum_{(\alpha \rho, \alpha' \rho')}\tilde{q}_{\rho \rho'}(\alpha, \alpha')q_{\rho \rho'}(\alpha, \alpha')
 +\frac{\alpha \beta}{2M^{2}}\sum_{\alpha \alpha' \rho}\tilde{R}_{\rho}(\alpha, \alpha')R_{\rho}(\alpha, \alpha')\notag \\
 &\hphantom{={}}
 -\frac{1}{\beta}\Bigl[\ln \Tr \exp \Bigl\{\frac{\beta}{M}\sum_{\alpha \rho}\xi \tilde{m}_{\rho}(\alpha)\sigma_{\rho}^{z}(\alpha) + \frac{\beta}{M}\sum_{\alpha \rho}\tilde{m}_{\rho}^{x}(\alpha)\sigma_{\rho}^{x}(\alpha)\notag \\
 &\hphantom{={}}
 +\frac{\alpha \beta^{2}}{2M^{2}} \sum_{(\alpha \rho, \alpha' \rho')}\tilde{q}_{\rho \rho'}(\alpha, \alpha')\sigma_{\rho}^{z}(\alpha)\sigma_{\rho'}^{z}(\alpha')
 +\frac{\alpha \beta^{2}}{2M^{2}} \sum_{(\alpha \alpha' \rho)}\tilde{R}_{\rho}(\alpha, \alpha')\sigma_{\rho}^{z}(\alpha)\sigma_{\rho}^{z}(\alpha')\Bigr\} \notag \\
 &\hphantom{={}}
 \times\prod_{\alpha \rho} \langle \sigma_{\rho}^{z}(\alpha) | \sigma_{\rho}^{x}(\alpha) \rangle
 \langle \sigma_{\rho}^{x}(\alpha) | \sigma_{\rho}^{z}(\alpha +1) \rangle \Bigr].\label{eq: Hopfield many pattern k body tilde f}
\end{align}

In a similar manner to the case of $k=2$, we look for the RS solution, and use the static ansatz.
The spin-product term in Eq.~\eqref{eq: Hopfield many pattern k body tilde f} is linearized by using a Gaussian integral.
Expanding the configurational term in Eq.~\eqref{eq: Hopfield many pattern k body tilde f} in powers of $n$, we have,
\begin{align}
 &n \biggl[\int Dz \ln \int Dw \Tr \exp \Bigl\{\frac{\beta}{M}\sum_{\alpha} \xi \tilde{m}\sigma^{z}(\alpha)
 +\frac{\beta}{M}\sum_{\alpha}\tilde{m}^{x}\sigma^{x}(\alpha)\notag \\
 &
 +\sqrt{\alpha \tilde{q}}\frac{\beta}{M}\sum_{\alpha}\sigma^{z}(\alpha)z + \sqrt{\alpha (\tilde{R}-\tilde{q})}\frac{\beta}{M}\sum_{\alpha}\sigma^{x}(\alpha)w\Bigr\}\notag \\
 &
 \times \prod_{\alpha}\langle \sigma^{z}(\alpha) | \sigma^{x}(\alpha) \rangle
 \langle \sigma^{x}(\alpha) | \sigma^{z}(\alpha +1) \rangle\biggr] + \mathrm{O}(n^{2}).
\end{align}
The inverse operation of the Trotter decomposition leads to
\begin{align}
 n \int Dz \ln \int Dw \, 2 \cosh \beta\sqrt{\Bigl(\tilde{m}+ \sqrt{\alpha \tilde{q}}z + \sqrt{\alpha (\tilde{R}-\tilde{q})}w\Bigr)^{2} + (\tilde{m}^{x})^{2}}
 + \mathrm{O}(n^{2}).
\end{align}
Using the replica trick, we finally obtain the following pseudo free energy:
\begin{align}
 f &= -s \lambda m^{k} + s(1-\lambda)(m^{x})^{2} - (1-s)m^{x}
 + \tilde{m}m + \tilde{m}^{x}m^{x} \notag \\
 &\hphantom{={}}
 +\frac{\alpha}{2}\beta (s \lambda)^{2} q^{k} - \frac{\alpha}{2}\beta (s \lambda)^{2}R^{k}
 -\frac{\alpha}{2}\beta\tilde{q}q + \frac{\alpha}{2}\beta \tilde{R}R \notag \\
 &\hphantom{={}}
 -\frac{1}{\beta}\int Dz \ln \int Dw \, 2 \cosh \beta\sqrt{\Bigl(\tilde{m}+ \sqrt{\alpha \tilde{q}}z + \sqrt{\alpha (\tilde{R}-\tilde{q})}w\Bigr)^{2} + (\tilde{m}^{x})^{2}}.\label{eq: pseudo free energy k>2}
\end{align}

The saddle-point conditions for the pseudo free energy~\eqref{eq: pseudo free energy k>2} yield the self-consistent equations.
Let 
\begin{align}
 g \equiv s \lambda k m^{k-1} + \sqrt{\alpha \tilde{q}}z + \sqrt{\alpha (\tilde{R}-\tilde{q})}w,\label{eq: g k>2}
\end{align}
and $u$ in the same way as in Eq.~\eqref{eq: def u}, and $Y$ as in Eq.~\eqref{eq: def Y}.
Then the self-consistent equations are given by Eqs.~\eqref{eq: m k=2}--\eqref{eq: m tilde k=2} with Eq.~\eqref{eq: g k>2} and
\begin{align}
 \tilde{q} &= (s \lambda)^{2}kq^{k-1},\\
 \tilde{R} &= (s \lambda)^{2}kR^{k-1}.
\end{align}

In the same way as the case of $k=2$, we find $R \ge q$.
If $R > q$, the free energy diverges in the limit $\beta \to \infty$.
Accordingly, $R$ must be equal to $q$.
It follows that $\tilde{R}=\tilde{q}$, so that the integrands in the self-consistent equations are independent of $w$.
Hence, the self-consistent equations in the low-temperature limit are
\begin{align}
 m &= \int Dz\,\frac{s \lambda (km^{k-1} + \sqrt{\alpha kq^{k-1}}z)}{\sqrt{(s \lambda [k m^{k-1} + \sqrt{\alpha k q^{k-1}}z])^{2}+(1-s-2s(1-\lambda)m^{x})^{2}}},\\
 m^{x} &= \int Dz\,\frac{1-s-2s(1-\lambda)m^{x}}{\sqrt{(s \lambda [km^{k-1} + \sqrt{\alpha k q^{k-1}}z])^{2}+(1-s-2s(1-\lambda)m^{x})^{2}}},\\
 q &= \int Dz\,\frac{(s \lambda [k m^{k-1} + \sqrt{\alpha kq^{k-1}}z])^{2}}{(s \lambda [km^{k-1} + \sqrt{\alpha kq^{k-1}}z])^{2}+(1-s-2s(1-\lambda)m^{x})^{2}}.
\end{align}
The factor $\beta (R-q)$ converges to
\begin{align}
 C &\equiv \lim_{\beta \to \infty} \beta (R-q) = \int Dz\,\frac{\{1-s-2s(1-\lambda)m^{x}\}^{2}}{\{(s \lambda [k m^{k-1} + \sqrt{\alpha k q^{k-1}}z])^{2}+(1-s-2s(1-\lambda)m^{x})^{2}\}^{3/2}}.
\end{align}
Since the factor $\beta (R^{k}-q^{k})$ converges to $Ckq^{k-1}$, the pseudo free energy in the low-temperature limit is
\begin{align}
 f &= s \lambda (k-1) m^{k} - s(1-\lambda)(m^{x})^{2} + \frac{\alpha}{2}k(k-1)(s \lambda)^{2}Cq^{k-1}
 \notag \\
 &\hphantom{={}}
 - \int Dz \sqrt{(s \lambda [k m^{k-1} + \sqrt{\alpha k q^{k-1}}z])^{2}+(1-s-2s(1-\lambda)m^{x})^{2}}.
\end{align}

\end{document}